# Long-Term Forecasts of Statewide Travel Demand Patterns Using Large-Scale Mobile Phone GPS Data: A Case Study of Indiana

**Rajat Verma**

PhD Candidate
Lyles School of Civil Engineering, Purdue University
550 W Stadium Ave, West Lafayette, IN 47907
Tel: 517-481-0901; Email: verma99@purdue.edu

**Eunhan Ka**

PhD Student
Lyles School of Civil Engineering, Purdue University
550 W Stadium Ave, West Lafayette, IN 47907
Tel: 765-586-8154; Email: kae@purdue.edu

**Satish V. Ukkusuri\***

Reilly Professor of Civil Engineering,
Lyles School of Civil Engineering, Purdue University
550 W Stadium Ave, West Lafayette, IN 47907
Tel: 650-454-0637; Email: sukkusur@purdue.edu

\* Corresponding author

**ABSTRACT**

The growth in availability of large-scale GPS mobility data from mobile devices has the potential to aid traditional travel demand models (TDMs) such as the four-step planning model, but those processing methods are not commonly used in practice. In this study, we show the application of trip generation and trip distribution modeling using GPS data from smartphones in the state of Indiana. This involves extracting trip segments from the data and inferring the phone users' home locations, adjusting for data representativeness, and using a data-driven travel time-based cost function for the trip distribution model. The trip generation and interchange patterns in the state are modeled for 2025, 2035, and 2045. Employment sectors like industry and retail are observed to influence trip making behavior more than other sectors. The travel growth is predicted to be mostly concentrated in the suburban regions, with a small decline in the urban cores. Further, although the majority of the growth in trip flows over the years is expected to come from the corridors between the major urban centers of the state, relative interzonal trip flow growth will likely be uniformly spread throughout the state. We also validate our results with the forecasts of two travel demand models, finding a difference of 5–15% in overall trip counts. Our GPS data-based demand model will contribute towards augmenting the conventional statewide travel demand model developed by the state and regional planning agencies.

## INTRODUCTION

Travel demand forecasting is one of the most common tasks in transportation planning. It involves prediction of trip frequency patterns and their impact on the transportation network, especially for highway and urban street networks. It helps in understanding how people move and make travel choices within a region, which in turn influences transportation infrastructure and policies (*1*). In the United States, travel demand forecasting is carried out routinely by the state departments of transportation (DOTs) as well as local and metropolitan planning agencies (*2*). Over the years, the state DOTs have largely converged on the methods for demand forecasting, though the rates of acceptance of newer methods differ greatly among them (*2*).

Conventionally, travel demand forecasting is performed with the help of the four-step planning model, which include (i) trip generation, (ii) trip distribution, (iii) mode choice assignment, and (iv) traffic assignment (*3*). The data used to generate estimates for the first three steps of this model generally come from travel surveys, where the subjects provide their travel choice preferences and typical trip making behavior. Similarly, newer methods such as activity-based planning also often require sophisticated travel survey data for generating reliable trip forecasts, sometimes even including Global Positioning System (GPS) trackers (*4*). However, conventional and GPS travel surveys have shortcomings such as high implementation cost, small sample size, and low temporal range (*5*, *6*).

In this context, the widespread use of GPS-enabled vehicles and mobile devices has given rise to huge amounts of passively collected geolocation data (*7*). These include call detail records (CDRs) from mobile phones, GPS traces from smartphones and vehicles, geotagged social media posts, smartcard transactions in public transit trips, among others. This kind of geolocation data offers scalability at low costs, large samples, and high spatiotemporal coverage and frequency, but suffers from a lack of sociodemographic data of the tracked individuals, sampling bias, and unknown trip characteristics (*8*, *9*).

There has been abundant research in using cellular geolocation data in the development of the traditional four-step planning model, including model components of all its four stages (*4*, *10*, *11*). However, most of the research pertains to CDRs (*8*, *12*). The methods used in the case of GPS data from mobile phones for modeling the four stages of the model are few and relatively less examined (*8*). Historically, this difference was driven by the abundance of CDR data from regular cell phones. The advent of large-scale GPS data from location-based services (LBS) on smartphones has now made it possible to use smartphone-based GPS geolocation data for this purpose. This is particularly relevant for aiding the development of large-scale travel demand models, such as megaregional or state traffic demand models (STDMs), which have long relied on traditional methods such as household travel surveys because of the absence of large-sample detailed travel behavior (*2*).

In this work, we focus on the first two steps of the four-step planning model, namely trip generation and trip distribution. Trip generation via GPS data involves inferring the trips made by the mobile device users and modeling their relationship with socioeconomic attributes (SEAs) of the traffic analysis zone (TAZ) of their home location estimated from their GPS traces. The forecasts of the SEAs are considered as inputs that are typically modeled based on demographic and economic growth models. The inferred trips are also used to compute the origin-destination matrix (ODM) of the target region. The developed trip generation model is used to predict the trip production and attraction patterns of the TAZs in the future, while a trip distribution model is used to predict the future trip interchanges between the TAZs based on these trips. In doing so, we show how large-scale mobile phone GPS data can be used to refine the four-step conventionally used in many STDMs, including using observed trip ODMs and interzonal costs in terms of travel times rather than Euclidean distances.

We apply this framework to Indiana statewide transportation planning, one of the many states that has so far relied on travel surveys for its STDM (*2*). We obtain the SEA data from the Indiana STDM (ISTDM) obtained from the Indiana Department of Transportation (INDOT) which contains the observed and predicted SEA values for four years – 2015, 2025, 2035, and 2045. We generate the ODMs of the state at the U.S. census-based TAZs for a typical weekday and weekend in these three future years (2025, 2035, 2045). The predicted travel patterns are analyzed and validated using a traditional four-step model for the Fort Wayne metropolitan area of Indiana.

The rest of the paper is organized as follows. First, we review key literature of travel demand modeling in the *State of Practice* section, including the methods proposed for large-scale mobility data from CDRs and GPS and the state of practice of STDMs in the United States. Then, we describe the socioeconomic data, methods for extracting trips from the GPS data, and the trip generation and distribution models used in the *Data and Methods* section. The results of the models and their predictions are analyzed in the *Results* section. The value and limitations of this framework are discussed in the *Conclusion* section.

## STATE OF PRACTICE

### Travel Demand Forecasting

Long-term travel demand forecasts assume significant importance, undergirding strategic planning and investment decision-making procedures (*13*). It has traditionally relied on data sources like household travel surveys which often require significant financial, human, and time investments for thorough data collection (*14*, *15*). The emergence of location-based data, in particular, given its ubiquity and substantial volume, has significantly bolstered our capacity to conduct expansive spatial and temporal pattern analyses within urban environment (*16*, *17*). This type of high-resolution data source presents new opportunities for comprehending and predicting mobility patterns, thereby refining the precision of travel demand forecasting.

### Four-Step Planning Model

The four-step planning model is a commonly used methodology in transportation planning, which involves four stages: trip generation, trip distribution, mode choice, and trip assignment. Trip generation, the first step in the four-step planning model, calculates the number of trips in a specific area based on socio-demographics, employment, and land use, using statistical models. Trip distribution, the second step, predicts the volume of trips between different origin-destination pairs. This involves modeling travel patterns and estimating trip interchanges ('flows') between zones. It is often based on the gravity model introduced by Wilson (1967) (*18*), which accounts for factors like distance and attractiveness of destinations. The third step, mode choice, determines the transport mode for each trip considering travel time, cost, convenience, and personal preferences. McFadden et al. (1973) (*19*) introduced the conditional logit model, which is widely used to model mode choice behavior. Finally, traffic assignment specifies the routes that travelers will take and the traffic volumes thus generated on each network link based on Wardrop's user equilibrium principle (*20*), which suggests travelers choose cost-minimizing routes.

### GPS Data in Transport Planning

Travel surveys are extensively utilized to develop travel demand models. The traditional approach for travel demand modelling relies on household travel surveys, which involves detailed trip and travelers' information. However, traditional survey methods have been plagued with limitations such as high cost, non-response rates, low sample sizes and data inaccuracies, making GPS technology a potential alternative to provide more reliable and precise data (*21*).

Chung and Shalaby (2005) proposed a trip reconstruction tool for GPS-based personal travel surveys, which extracts travel information such as trip purpose, mode of transportation, and destination from GPS data (*22*). Bohte et al. (2009) presented a methodology for deriving and validating trip purposes and travel modes from multi-day GPS-based travel surveys (*23*). The authors posited that GPS data can yield more accurate and comprehensive travel data than traditional survey methods, but reliable methods for processing and analyzing the data are crucial. Hong et al. (2021) provided insights on data quality from a large-scale application of smartphone-based travel survey technology in the Phoenix metropolitan area (*24*). They showed the importance of data completeness, accuracy, and representativeness, and highlighted challenges associated with smartphone-based travel surveys.

Based on previous studies, we summarize advantages for using GPS data in transport planning as (i) comprehensive data with substantial representation of the population (*25*), (ii) availability at large scales for relatively low cost (*26*), and (iii) sufficient temporal variability and spatial density so as to infer details such as trip chains, trip purpose, and travel modes with reasonable confidence (*27–29*). However, GPS data also bear several limitations, such as (i) data processing challenges, including the need for carefully choosing the methods and their parameters and computing resources (*28*, *30*) and (ii) validation with more direct methods such as travel surveys (*26*).

Though GPS data are increasingly used in large-scale travel demand models, such as STDMs, this effort is not evenly distributed across the states. Our research is poised to illustrate the incorporation of GPS data in travel demand prediction, and advancing the field of large-scale geolocation data analysis for transportation planning.

**DATA AND METHODS**

The central idea in this study is to use smartphone GPS data to estimate current travel patterns and use relevant socioeconomic attributes in the current and the future years to estimate the travel patterns (in the form of ODMs). This process is illustrated in Figure 1 and described in the following sections.

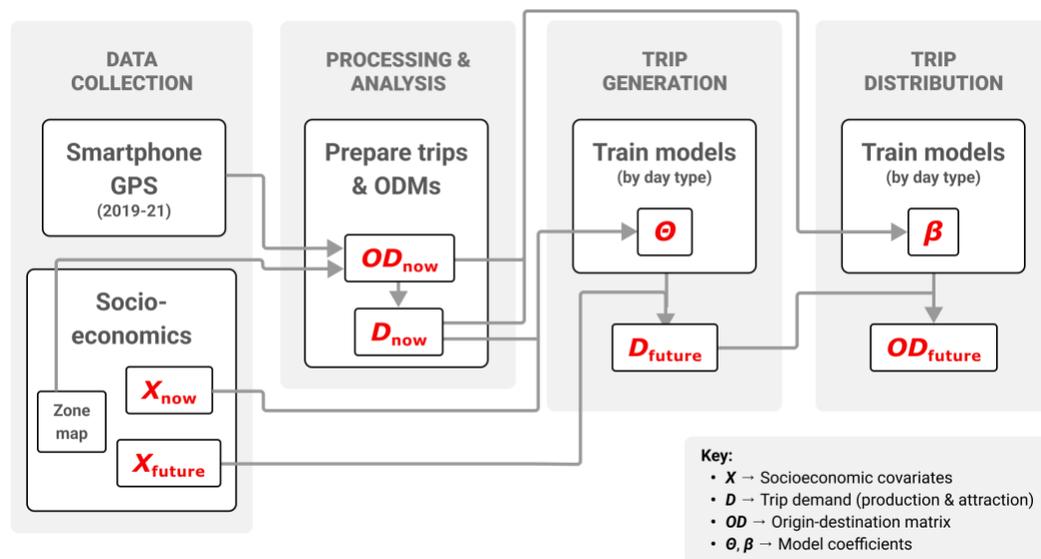

**Figure 1: Overall framework for trip forecasting using mobile phone data.**

After the smartphone data are collected, they are first cleaned and are then used to estimate the device owners' home zone and identify trip segments, where each segment represents the reconstruction of a person's trip by any single travel mode. The origins and destinations of these trips are used to generate

ODMs for different base years, segregated by day type. These constitute the current ODMs, called $OD_{now}$. By aggregating the trip count by origin and destination zones using ISTDM TAZ classification, the current trip attraction and generation are computed, referred to as $D_{now}$ in Figure 1. Patterns from these trips are visualized. Then, the current socioeconomic data attributes of ISTDM zones, $X_{now}$, are used to fit a set of trip generation models by day type.

**Socioeconomic Attributes**

The Indiana Statewide Travel Demand Model (ISTDM) (*31*) is used to forecast future travel patterns in Indiana. It is a comprehensive modeling system that integrates various data sources, such as demographic, land use, and transportation data, to develop an understanding of travel patterns in Indiana. This ISTDM is based on a four-step planning model. The demographic and economic factors are provided for 4,703 TAZs within Indiana. The ISTDM covariate values are normalized with respect to the population to make them comparable across the different regions of comparison (e.g., counties or TAZs) and converted into 12 explanatory variables, including indicators of population, income, and employment in different industries.

**GPS Data Processing and Representativeness**

The GPS data used in this study are obtained from a data vendor. Each record of the original pings is given by a 4-tuple: (longitude, $x$, latitude, $y$, UNIX-timestamp, $t$, GPS spatial error, $\varepsilon$). The records are first filtered for high GPS accuracy of $\varepsilon > 50$ m. Next, only high-quality device users are filtered so that there is reasonable confidence in the subsequent steps using data quality-quantity tradeoff assessment. To achieve this, a double temporal frequency matrix is constructed for each base year (2019, 2020, and 2021), based on the work of Scheinder et al. (2013) (*32*). An example frequency matrix is shown in Figure 2A which shows the distribution of the number of filtered users in Indiana in 2021 for different combinations of frequency thresholds. A threshold of 10 half-hour bins and 1 day is chosen to filter the ping records, as highlighted in panel A. This means that the filtered users have at least one ping in at least 10 of the 48 half-hour periods in at least one day. This process retains only 14% of the total users (A1), though these are high-quality users that contain 94% of all the pings (A2).

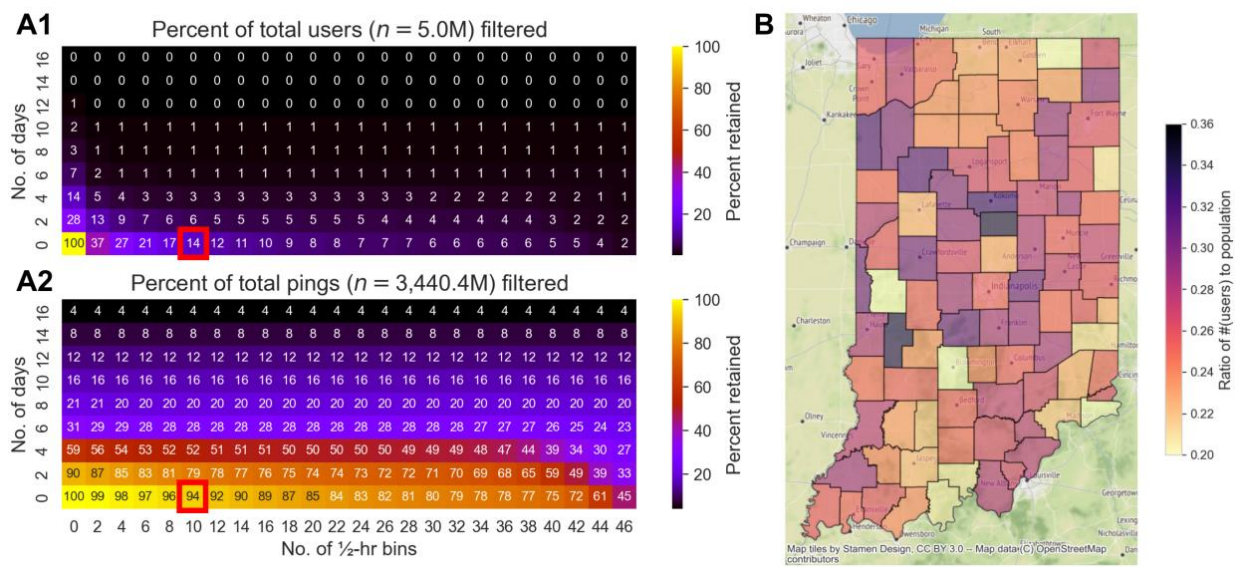

**Figure 2: Quality of the GPS data. (A) Data quality-quantity tradeoff matrices for the 2021 geolocation data. (B) Data representativeness at the county level.**

Once the data are filtered, it is important to compute their representativeness to see if there are any specific regions which show problematic representation of or bias in population in the dataset (*33*). It is defined as the ratio of the number of users whose homes are detected in a region to that region's population as of the 5-year estimate of the American Community Survey of 2020. The users' homes are detected using Yabe et al. (2022)'s method (*12*) that uses mean shift clustering to detect the most commonly visited place during nighttime (9 PM – 6 AM). The representativeness figures obtained in this way are shown in Figure 2B. At the county level, the unfiltered dataset includes information about 20 to 40% of the population, which is sufficient for most of the analysis even with considerable filtering (*34*).

**Trip Segmentation and OD Matrix Preparation**

We use the rule-based stay point detection method used in Li et al. (2008) (*35*) to detect trip endpoints. The stay points are chosen as centroids of the largest regions having two pings at most 100 m apart and within 10 minutes of each other (*36*). Once the trip segments are generated, their endpoints (origin and destination) are used to compute the total number of trips to and from different TAZs. These are then upscaled using the representativeness figures as explained in the previous section.

**Trip Generation Model**

A linear regression-based trip generation model is developed in this study to predict the production and attraction of home-based and work-based trips. To compute the attractiveness of the TAZs, the socioeconomic and land use indicators of the zones are assessed. The predictor variables are computed from the measures available in the ISTDM shapefiles for the four study years – 2015, 2025, 2035, and 2045.

These variables are computed so that they have minimal correlation among themselves. It can be seen that population density is highly correlated to the overall population of the TAZs ($\rho = 0.61$). All the other variable pairs have a correlation with a magnitude less than 0.5 and are thus considered safe for inclusion in the subsequent trip generation model.

**Trip Distribution Model**

Gravity model is a traditional trip distribution model which assumes a negative exponential dependence of trip interchange probability on the impedance between two zones (*37*). In this task, a variant of the standard gravity model is used, given by the following equation:

$$\widehat{N}_{ij}(\mathbf{P}, \mathbf{A}, \mathbf{D} \mid \beta) = \frac{A_j D_{ij}^{-\beta}}{\sum_{k=1}^{n} A_k D_{ik}^{-\beta}} \cdot P_i \quad \forall\, i, j \in 1{:}n \tag{1}$$

Here, $\widehat{N}_{ij}$ is the number of predicted trips between zones $i$ and $j$, $P_i$ is the number of trips produced in zone $i$, $A_j$ is the number of trips attracted by zone $j$, $D_{ij}$ is the interzonal cost function, and $n$ is the total number of zones. $\beta$, the only model parameter, is called the gravity exponent.

The computation of the cost function is contextualized to suit the given geolocation data in this project. Instead of using Euclidean distances between two zones (or a nonvariant equivalent conversion to travel time) as the measure of friction factor, we use the mean or median of the distance or travel time values of all the trips made in the study period between any two zone pairs. This method is better than the static conversion in that it accounts for the actual path of the trip along the roadway instead of the as-the-crow-flies path and also the temporal heterogeneity in trip-making behavior, such as weekdays versus weekends and peak and off-peak hours.

Once the data for the gravity model are obtained, the only parameter of the model – the gravity exponent, $\beta$, is calibrated using line search optimization. The parameter test range $[\beta_0, \beta_1]$ is set to $[0.5, 3.0]$ based on the accepted range in the literature (*38*, *39*).

## RESULTS

### Trip Generation

Trip generation models for trip production and attraction are developed as a set of linear regression models for a typical weekday and weekend at different spatial scales for different years. These models primarily make use of socioeconomic characteristics that have already been predicted for the future years in the ISTDM. The correlation half-matrices for these variables at the county and TAZ-level are shown in Figure 3. Except for total population and population density, the descriptive variables do not have significant correlation since the magnitudes of correlation coefficients are less than 0.5. This correlation is particularly high in the case of TAZs because unlike the Census-delineated regions, which focus on creating regions based on the uniformity of population distribution, TAZs typically focus on creating more uniformly distributed land areas. This results in large TAZs even in densely populated regions, causing a strong correlation between total population and population density. To avoid the issue of multicollinearity in the subsequent models, we did not include population density in the models.

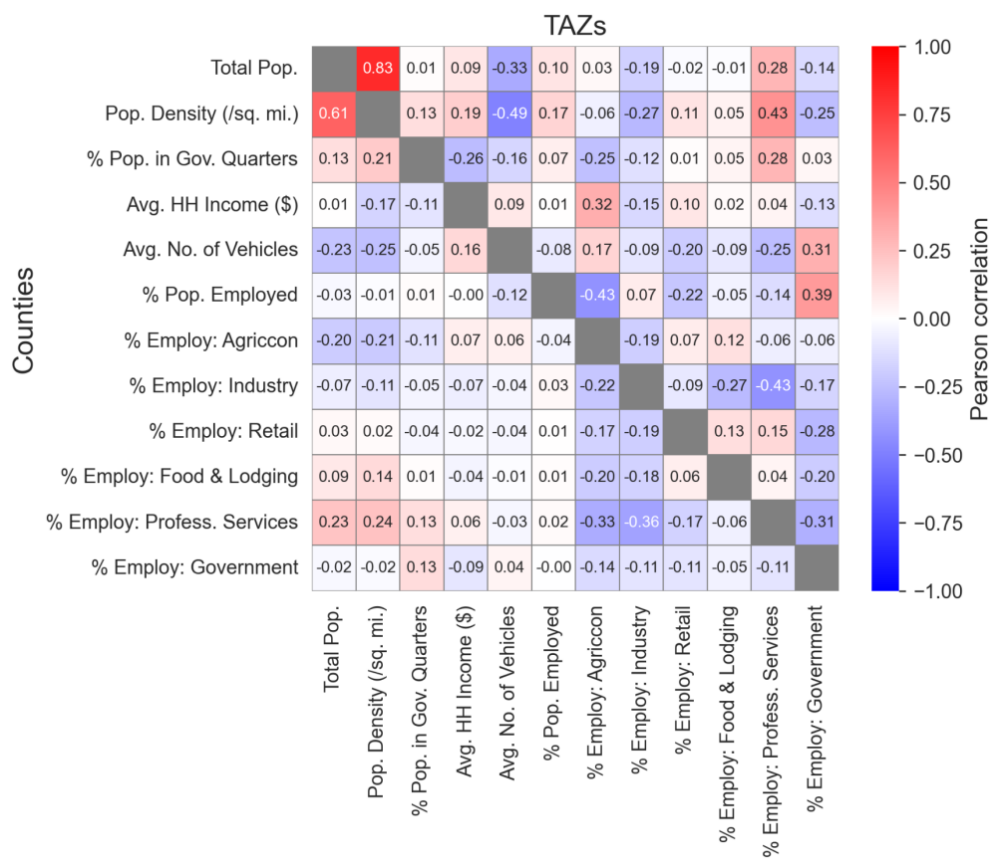

**Figure 3: Pearson correlation between the covariates used in the trip generation model at two scales: traffic analysis zones (upper right triangle) and county (lower left triangle).**

The models for weekday and weekend trip generation are based on trip data for 2021. This is because the models based on the 2019 data have low fitness ($R^2$) values and there might be possible confounding with the 2020 data because of partial substantial mobility pattern disruptions during the month of March due to COVID-19 (*40*).

The coefficients of the covariates and the model fitness statistics for the four finalized models are shown in Table 1. Across all the models, it is clear that population is one of the most important predictive factors of trip counts. This is consistent with many trip generation models (*41*). Also, all the other covariates have a strong positive impact on trip production and generation at the 95% confidence level. Some of the industry-specific employment covariates such as percent employment in industry (`INDUST`), retail (`RETAIL`), food and lodging (`FOODLD`), and professional services (`PROSRV`) are found to be significant indicators of trip count, while others such as the percent employment in agriculture (`AGCON`), government (`GOVNMT`), and other services (`OTHSRV`) are not found to be significant, in alignment with the findings of other research (*42*).

**Table 1: Coefficients of the finalized trip generation models for the total number of population - adjusted trips on a typical weekday/weekend.**

| Model→<br>Covariate ↓ | Production Weekday | Attraction Weekday | Production Weekend | Attraction Weekend |
|---|---|---|---|---|
| Constant | -526.146*** | -516.757*** | -374.123*** | -369.917*** |
|  | (88.72) | (91.62) | (66.50) | (72.78) |
| Total Population | 0.483*** | 0.482*** | 0.356*** | 0.361*** |
|  | (0.00) | (0.00) | (0.00) | (0.00) |
| % Pop. in Government Quarters | 1344.826*** | 1411.755*** | 475.288*** | 611.110*** |
|  | (150.08) | (154.98) | (112.48) | (123.11) |
| Avg. Household Income ($) | 0.003*** | 0.002*** | 0.003*** | 0.002*** |
|  | (0.00) | (0.00) | (0.00) | (0.00) |
| Avg. No. of Vehicles | 214.119** | 212.882** | 185.729*** | 188.649*** |
|  | (84.64) | (87.40) | (63.44) | (69.43) |
| % Pop. Employed | 3.406*** | 3.600*** | 1.764*** | 2.006*** |
|  | (0.55) | (0.57) | (0.41) | (0.45) |
| % Employed: Industry | 366.771*** | 371.531*** | 129.971*** | 137.377** |
|  | (66.24) | (68.40) | (49.65) | (54.34) |
| % Employed: Retail | 637.368*** | 673.878*** | 514.551*** | 598.531*** |
|  | (82.63) | (85.33) | (61.93) | (67.78) |
| % Employed: Food & Lodging | 896.288*** | 945.145*** | 789.855*** | 884.140*** |
|  | (107.66) | (111.18) | (80.69) | (88.32) |
| % Employed: Professional Services | 381.680*** | 390.159*** | 123.195** | 124.241** |
|  | (68.99) | (71.25) | (51.71) | (56.60) |
| Adjusted $R^2$ | 0.770 | 0.758 | 0.761 | 0.733 |
| F-statistic | 1357.0 | 1272.0 | 1291.0 | 1117.0 |
| (Probability) | (0.0) | (0.0) | (0.0) | (0.0) |

The asterisks next to the coefficients denote the p-value level: *: $p < 0.1$, **: $p < 0.05$, ***: $p < 0.01$. The values in the parentheses denote the standard error of the covariates.

*Predicted Trip Generation*

The finalized linear regression models, described in Table 1, are used to predict the trip generation and attraction for three future and one past year in the ISTDM dataset – 2015, 2025, 2035, and 2045. For these years, the total attraction and production by weekday and weekend are shown in Figure 4(A). Across Indiana, the models predict a total of 24.9 and 20.3 million trips on a typical weekday and weekend in 2015 respectively. According to the models, these numbers are expected to grow at a steady decadal growth rate of 3.9 to 4.6 percent (panel B) over the three decades, leading to a total of 28.4 and 23.7 million trips by 2045. Notably, the differences in total trip attraction and production are not substantial.

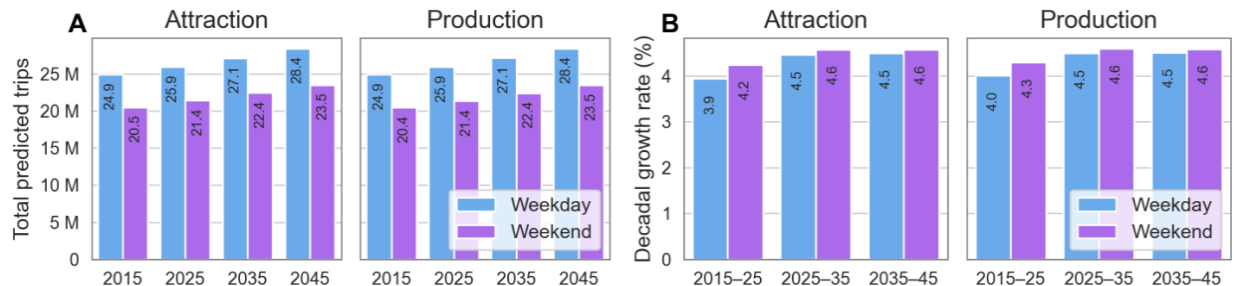

**Figure 4: Predicted trip counts at the TAZ level for the study years, showing (A) total produced and attracted trips on a typical weekday and weekend, and (B) the decadal growth rates in the trip counts.**

The trip generation resulting in values and growth also show considerable spatial heterogeneity over the years. Figure 5 shows the trip production at the TAZ level, i.e., the number of predicted trips produced in each TAZ. The broad patterns of trip density remain largely the same between 2015 (panel A) and 2045 (panel B) in that a huge proportion of the trips are centered in the major urban regions, primarily the Indianapolis and Chicago-Gary metropolitan statistical areas (MSAs). These are consistent with the population density distribution of the state at the TAZ level.

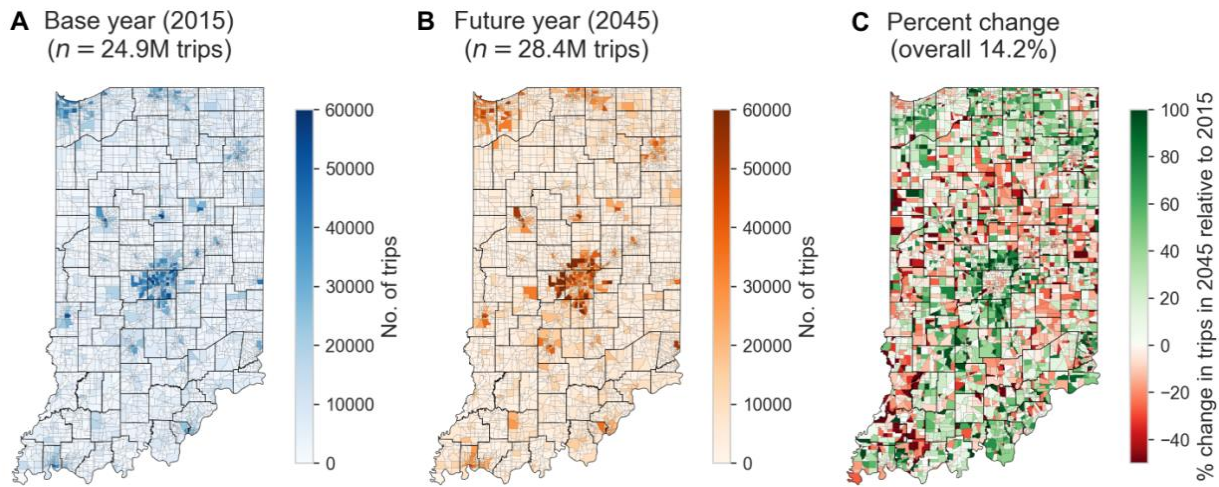

**Figure 5: Comparison of weekday trip production in TAZs between (A) 2015 and (B) 2045. Panel (C) shows the percent growth in trip counts.**

However, there are significant differences even at the TAZ level. This is shown explicitly in panel C which highlights the zones by their trip production growth rate between these three decades. The green regions that show positive growth are primarily centered around the northern belt from Chicago to Fort Wayne as well as in the suburbs of Indianapolis. Particularly, the regions close to Indianapolis core but outside its parent Marion County show some of the highest growth during this period. This intuitively makes sense since cities generally grow outward in the suburbs (*43*). However, it is not explicitly checked if this growth occurs due to the increase in population of the suburbs or because of the change in the employment and sectoral distribution of the jobs.

Conversely, the zones inside Marion County broadly show a small decline in trip production, suggesting a general decline in the level of activity in the region. Similar patterns are also seen in the western region of the state, close to the border with Illinois, and in the belt between the northern corridor around Fort Wayne and the center around Indianapolis.

*Regions with the Greatest and Least Growth*

Figure 6 shows the top and bottom performing counties in Indiana by the percent growth rate of predicted trips between 2015 and 2045. For both trip production and attraction, the top 5 counties with the greatest growth are in the Indianapolis MSA, with upwards of 40% growth over the three decades. Notably, Marion County, which seats the core of the city, is associated with a small decline in the trip count. This indicates a substantial shift of population and opportunities from the city core to the suburbs of the largest city of Indiana. This is consistent with the trend of other major cities in the United States over the years (*43*).

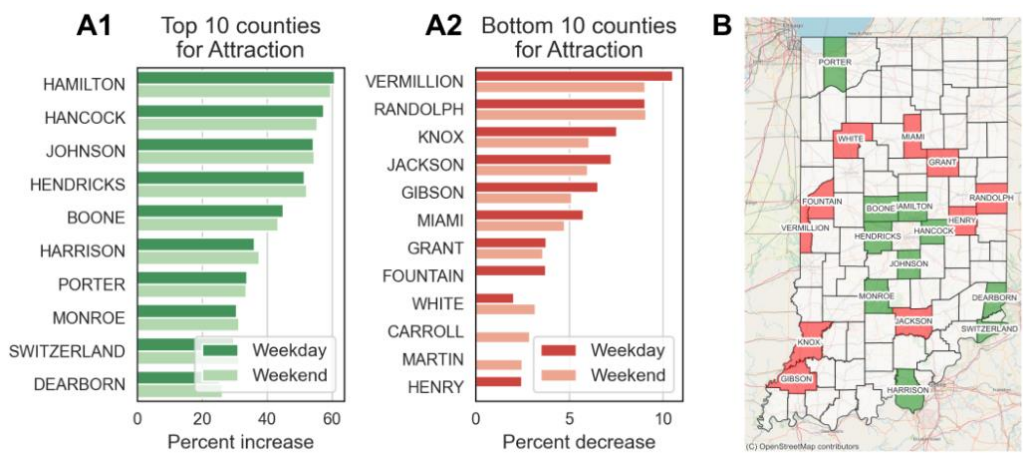

**Figure 6: Top and bottom 10 counties in Indiana by the percent change in predicted trips between 2015 and 2045 for trip production and attraction on a typical weekday and weekend: (A) Values and (B) Map.**

**Trip distribution**

*Model Calibration*

The gravity models for the trip distribution modeling are calibrated using line search optimization because of the presence of ground truth data and there being only one parameter in each model – the gravity exponent, $\beta$. As explained in the Methodology section, the model is calibrated for each of the three base years (2019, 2020, and 2021), both day types (weekday and weekend), and additionally four cost functions (median/mean of travel time/trip length). The test range for the gravity exponent is taken as 0.1 to 3.0. The results of the line search optimization, observed in terms of mean squared error (MSE) of the fitted model's trip count from the ground truth data of 2021.

It is generally seen that the travel time MSE curves of travel time aggregations (median and mean) are continuously increasing and thus have the best fitted $\beta$ values at the range start (i.e., $\beta = 0.1$), whereas the ones for distance (trip length)-based aggregations have distinct minima at around $\beta = 1.5$. Based on the preference of distance over travel time and the general preference of the median over the mean in the case of skewed distributions, the models with median distance are chosen. This led to a value of $\hat{\beta} = 1.6$ for a typical weekday and $\hat{\beta} = 1.4$ for a weekend. This higher value of the gravity exponent for weekdays compared to weekends is in alignment with the literature (*44*) as it indicates the impact of travel time on commute trips on weekdays compared to more leisure trips on weekends.

*Change of Trip Flows*

The inter-county flows predicted from the calibrated model are shown in Figure 7. Most of the heavy traffic corridors are predicted to be between Indianapolis and the major urban centers across the state, particularly the Chicago-Gary MSA and Fort Wayne (panel A). These major corridors remain uncontested across the years and contribute to the greatest growth in flows between 2015 and 2045 (panel B). However, when taking into factor the growth rate (i.e., normalizing with respect to base year's flows), the distribution changes completely (panel C). There seems to be substantial heterogeneity in the growth rate across different OD pairs, but most of these are inconsequential in absolute numbers, as seen in the differences in the flow maps in panels B and C.

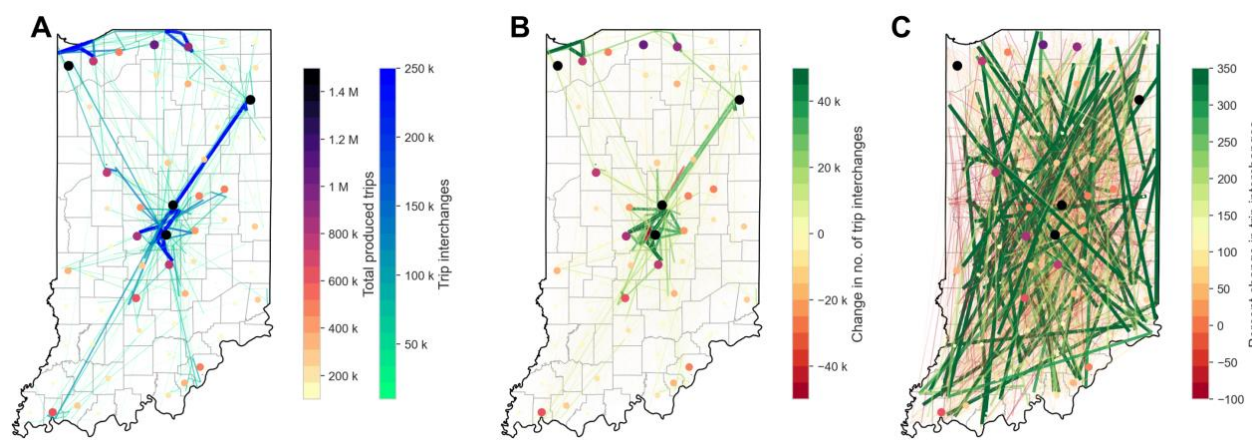

**Figure 7: Maps of the predicted weekday inter-county flows and changes over the years: (A) Flows (trip interchanges) in 2045, (B) change in flows between 2015 and 2045, and (C) percent change in flows between 2015 and 2045. The total county trip production of 2045 is shown in each panel by colored dots, with darker shades representing busy counties (legend in panel A).**

**Comparison and Validation**

*Comparison with Indiana STDM*

To strengthen the validity of our results from the GPS data, we compared the trip generation predictions with the ISTDM's trip demand figures. We obtained the ISTDM's ODM for 2015 from the Indiana DOT containing a symmetric OD matrix for the TAZs for automobiles. The relationship between the trip production from ISTDM and our model are shown in Figure 8A at the TAZ-level. Their spatial relationship is shown in panel B.

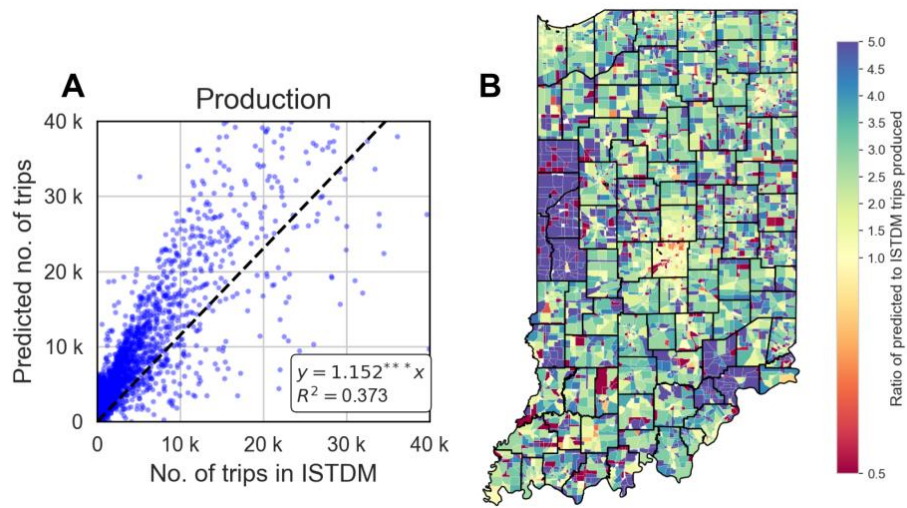

**Figure 8: Comparison of trip production predicted by the trip generation model with ISTDM trips for 2015. (A) Scatter plot showing a linear model fit, with statistically significant slope at 95% confidence level (B) Map showing the ratio of predicted to ISTDM values for each TAZ, with black lines showing the county borders.**

Through this figure, it is clear that the geolocation-based predictions are strongly correlated ($\rho = 0.82$) with the ISTDM figures. However, they are also observed to be consistently overestimated compared to ISTDM's predictions. The fit of a univariate linear regression shows a slope of 1.152, indicating the GPS data-based predictions being overestimated by about 15%. It should be noted, however, that the underlying trip count for the GPS-based model depends on several parameters that may be tuned specifically for a target region so as to achieve a target prediction count. This includes the clustering parameters (such as the cluster radius or the maximum inter-cluster distance) (*45*), the duration of the stay region, or a minimum number of pings or data quality for the extracted trips.

In addition to the observation of a slight overestimation of the overall trip count from the GPS-based trip generation model, it is also observed that these values are also spatially heterogeneously distributed. In Figure 8B, it can be seen that there is a much higher number of trips predicted by the GPS-based model in the western and the southeastern portions of the state, which tend to be rural. It might be attributable to inadequate representativeness figures of these regions which highly skew the results. Notably, the predicted figures match very closely with the ISTDM figures (ratio close to unity) in Marion County, where there are sufficient mobile phone users to be adequately represented in the GPS dataset. This points out to the general observation that larger and more representative GPS data tend to show better results in predicting human mobility patterns (*46, 47*).

*Validation with Fort Wayne Model*

To strengthen the validity of our results from the GPS data, we compare the results of the developed trip generation models with the results of the Fort Wayne Travel Demand Model (FWTDM) (*48*) whose planning region and period closely resemble those used in this study. The Fort Wayne transportation plan of 2040 was developed on the basis of 2015 data. Its planning area includes most of Allen County, which includes most of the Fort Wayne city boundary, and parts of nearby rural counties – Whitley and Huntington, shown in Figure 9A. Trip productions and attractions were developed for different trip purposes. To perform a fair comparison of the FWTDM with our model's results in the absence of a boundary GIS layer, we include all of Allen County and compare linearly interpolated figures between 2015 and 2045.

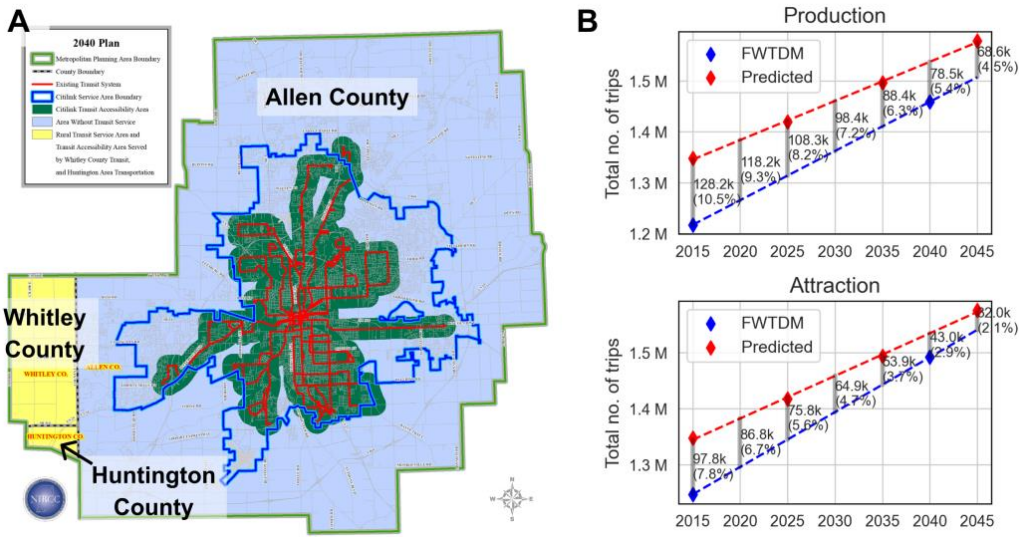

**Figure 9: Comparison of predicted trip generation values with the values from the FWTDM. (A) Map of the planning area in FWTDM showing parts of the included counties, taken from (*48*). (B) Year-wise differences in the trip counts.**

It can be seen in Figure 9B that our model consistently overestimates the number of trips by about 5–10% compared to the FWTDM. This is similar to the results of the previous sections and might be attributed to the choice of data filtering and trip segmentation parameters. Further, our model also predicts a smaller growth in overall travel in the Fort Wayne area over the coming decades compared to the FWTDM, though the difference is small. For example, in 2015, our model overestimates ≈97,800 trips, but this difference reduces to ≈32,000 by 2045. This might be caused due to a huge number of differences in the two models, such as different SEA data and region boundary. However, we deem this difference small enough for calling our predictions reasonable.

**CONCLUSION**

This study extends the traditional four-step model for large-scale travel demand forecasting using GPS geolocation data from mobile phones, such as for statewide travel demand models (STDMs). Though multiple U.S. states have traditionally used household travel surveys to obtain trip making behavior, the reluctance to integrate large-scale geolocation data analysis into this mobility assessment persists. In this study, we show methods for cleaning and processing the GPS data to predict current trip patterns and use them in aiding the four-step model. In doing so, we argue that the trips obtained from the GPS data provide a better representation of the population at large scale, the data quality can be sufficient to make accurate current trip counts, and use the estimated trip durations as the cost functions for the trip distribution model.

Our findings from this case study are multifaceted and overarching. We find several factors, particularly total population and percent employment in specific sectors like industry and retail, to be major contributors to both trip production and attraction. In the trip generation predictions, it is clear that the growth of movement in the state is expected to be at a nearly constant rate albeit with substantial spatial heterogeneity. Travel within the cores of the largest urban centers is predicted to decline mildly, but travel will increase the most to and from their suburbs. For example, all the 5 highest growing counties lie in the Indianapolis while its core Marion County is expected to have a small decline in overall trip production and generation. The same pattern is found in the inter-city flows, where the suburban cities are predicted to have the greatest growth in absolute counts, but this pattern vanishes when compared in terms of relative growth. We anticipate other U.S. states with similar geographical and socioeconomic patterns should follow similar patterns and interpretations.

We also partially validate our results with the forecasts of other models. Particularly, we compare our predictions with those of both Indiana's ISTDMs as well as Fort Wayne's regional TDM. In both the cases, our model shows an overprediction of the trips by 5–15%. This difference is the least in the main urban areas which have on average higher representation in the GPS data. We argue that the methods (and their parameters) used to identify the trips using the GPS data play a significant role in influencing the overall trip counts.

We also acknowledge the limitations of this study and directions for future research. These include the need of a detailed assessment of the processing for large geolocation data and their assumptions and a pilot study in a small area to compare and validate the GPS data-based estimates of the current travel patterns with a GPS travel survey.

## ACKNOWLEDGMENTS

This study was conducted as part of project number SPR 4608 of the Joint Transportation Research Program at Purdue University in collaboration with the Indiana Department of Transportation.

## AUTHOR CONTRIBUTION

The authors confirm contribution to the paper as follows: study conception and design: Ukkusuri, S.V.; literature review: Ka, E.; data collection and model training: Verma, R.; draft manuscript preparation: Verma, R., Ka, E., and Ukkusuri, S.V. All authors reviewed the results and approved the final version of the manuscript.

## REFERENCES

1. Çolak, S., L. P. Alexander, B. G. Alvim, S. R. Mehndiratta, and M. C. González. Analyzing Cell Phone Location Data for Urban Travel: Current Methods, Limitations, and Opportunities. *Transportation Research Record*, Vol. 2526, 2015, pp. 126–135. https://doi.org/10.3141/2526-14.

2. Donnelly, R., and R. Moeckel. *Statewide and Megaregional Travel Forecasting Models: Freight and Passenger*. The National Academies Press, Washington D.C., 2017.

3. McNally, M. G. The Four-Step Model. In *Handbook of transport modelling*, Emerald Group Publishing Limited, pp. 35–53.

4. Shen, L., and P. R. Stopher. Review of GPS Travel Survey and GPS Data-Processing Methods. Routledge, 34, 3, 2014, pp. 316–334.

5. Nitsche, P., P. Widhalm, S. Breuss, N. Brändle, and P. Maurer. Supporting Large-Scale Travel Surveys with Smartphones–A Practical Approach. *Transportation Research Part C: Emerging Technologies*, Vol. 43, 2014, pp. 212–221.

6. Cools, M., E. Moons, and G. Wets. Assessing the Quality of Origin–Destination Matrices Derived from Activity Travel Surveys: Results from a Monte Carlo Experiment. *Transportation research record*, Vol. 2183, No. 1, 2010, pp. 49–59.

7. Toole, J. L., S. Colak, B. Sturt, L. P. Alexander, A. Evsukoff, and M. C. González. The Path Most Traveled: Travel Demand Estimation Using Big Data Resources. *Transportation Research Part C: Emerging Technologies*, Vol. 58, 2015, pp. 162–177. https://doi.org/10.1016/j.trc.2015.04.022.

8. Wang, Z., S. Y. He, and Y. Leung. Applying Mobile Phone Data to Travel Behaviour Research: A Literature Review. *Travel Behaviour and Society*, Vol. 11, 2018, pp. 141–155. https://doi.org/10.1016/j.tbs.2017.02.005.

9. Jiang, S., J. Ferreira, and M. C. Gonzalez. Activity-Based Human Mobility Patterns Inferred from Mobile Phone Data: A Case Study of Singapore. *IEEE Transactions on Big Data*, Vol. 3, No. 2, 2016, pp. 208–219. https://doi.org/10.1109/tbdata.2016.2631141.

10. Çolak, S., L. P. Alexander, B. G. Alvim, S. R. Mehndiratta, and M. C. González. Analyzing Cell Phone Location Data for Urban Travel: Current Methods, Limitations, and Opportunities. *Transportation Research Record*, Vol. 2526, 2015, pp. 126–135. https://doi.org/10.3141/2526-14.

11. Toole, J. L., S. Colak, B. Sturt, L. P. Alexander, A. Evsukoff, and M. C. González. The Path Most Traveled: Travel Demand Estimation Using Big Data Resources. *Transportation Research Part C: Emerging Technologies*, Vol. 58, 2015, pp. 162–177. https://doi.org/10.1016/j.trc.2015.04.022.

12. Yabe, T., N. K. W. Jones, P. S. C. Rao, M. C. Gonzalez, and S. V. Ukkusuri. Mobile Phone Location Data for Disasters: A Review from Natural Hazards and Epidemics. *Computers, Environment and Urban Systems*, Vol. 94, 2022, p. 101777. https://doi.org/10.1016/J.COMPENVURBSYS.2022.101777.

13. He, Q.-C., and T. Hong. Integrated Facility Location and Production Scheduling in Multi-Generation Energy Systems. *Operations Research Letters*, Vol. 46, No. 1, 2018, pp. 153–157.

14. Cools, M., E. Moons, and G. Wets. Assessing the Quality of Origin–Destination Matrices Derived from Activity Travel Surveys: Results from a Monte Carlo Experiment. *Transportation research record*, Vol. 2183, No. 1, 2010, pp. 49–59.

15. Cheng, Z., S. Jian, T. H. Rashidi, M. Maghrebi, and S. T. Waller. Integrating Household Travel Survey and Social Media Data to Improve the Quality of Od Matrix: A Comparative Case Study. *IEEE Transactions on Intelligent Transportation Systems*, Vol. 21, No. 6, 2020, pp. 2628–2636.

16. Hasan, S., and S. V Ukkusuri. Urban Activity Pattern Classification Using Topic Models from Online Geo-Location Data. *Transportation Research Part C: Emerging Technologies*, Vol. 44, 2014, pp. 363–381.

17. Wang, Z., S. Wang, and H. Lian. A Route-Planning Method for Long-Distance Commuter Express Bus Service Based on OD Estimation from Mobile Phone Location Data: The Case of the Changping Corridor in Beijing. *Public Transport*, Vol. 13, 2021, pp. 101–125.

18. Wilson, A. G. A Statistical Theory of Spatial Distribution Models. *Transportation Research*, Vol. 1, No. 3, 1967, pp. 253–269. https://doi.org/https://doi.org/10.1016/0041-1647(67)90035-4.

19. McFadden, D., and others. Conditional Logit Analysis of Qualitative Choice Behavior. 1973.

20. Wardrop, J. G. Road Paper. Some Theoretical Aspects of Road Traffic Research. *Proceedings of the institution of civil engineers*, Vol. 1, No. 3, 1952, pp. 325–362.

21. Sarmiento, I., C. González-Calderón, J. Córdoba, and C. Díaz. Important Aspects to Consider for Household Travel Surveys in Developing Countries. *Transportation Research Record*, Vol. 2394, No. 1, 2013, pp. 128–136. https://doi.org/10.3141/2394-16.

Verma, Ka, and Ukkusuri 17

22. Chung, E.-H., and A. Shalaby. A Trip Reconstruction Tool for GPS-Based Personal Travel Surveys. *Transportation Planning and Technology*, Vol. 28, No. 5, 2005, pp. 381–401.

23. Bohte, W., and K. Maat. Deriving and Validating Trip Purposes and Travel Modes for Multi-Day GPS-Based Travel Surveys: A Large-Scale Application in the Netherlands. *Transportation Research Part C: Emerging Technologies*, Vol. 17, No. 3, 2009, pp. 285–297.

24. Hong, S., F. Zhao, V. Livshits, S. Gershenfeld, J. Santos, and M. Ben-Akiva. Insights on Data Quality from a Large-Scale Application of Smartphone-Based Travel Survey Technology in the Phoenix Metropolitan Area, Arizona, USA. *Transportation Research Part A: Policy and Practice*, Vol. 154, 2021, pp. 413–429.

25. Stopher, P., C. FitzGerald, and J. Zhang. Search for a Global Positioning System Device to Measure Person Travel. *Transportation Research Part C: Emerging Technologies*, Vol. 16, No. 3, 2008, pp. 350–369.

26. Hong, S., F. Zhao, V. Livshits, S. Gershenfeld, J. Santos, and M. Ben-Akiva. Insights on Data Quality from a Large-Scale Application of Smartphone-Based Travel Survey Technology in the Phoenix Metropolitan Area, Arizona, USA. *Transportation Research Part A: Policy and Practice*, Vol. 154, 2021, pp. 413–429.

27. Zou, X., S. Zhang, C. Zhang, J. J. Q. Yu, and E. Chung. Long-Term Origin-Destination Demand Prediction With Graph Deep Learning. *IEEE Transactions on Big Data*, Vol. 8, No. 6, 2022, pp. 1481–1495. https://doi.org/10.1109/TBDATA.2021.3063553.

28. Bohte, W., and K. Maat. Deriving and Validating Trip Purposes and Travel Modes for Multi-Day GPS-Based Travel Surveys: A Large-Scale Application in the Netherlands. *Transportation Research Part C: Emerging Technologies*, Vol. 17, No. 3, 2009, pp. 285–297.

29. Shen, L., and P. R. Stopher. Review of GPS Travel Survey and GPS Data-Processing Methods. *Transport Reviews*. 3. Volume 34, 316–334.

30. Chung, E.-H., and A. Shalaby. A Trip Reconstruction Tool for GPS-Based Personal Travel Surveys. *Transportation Planning and Technology*, Vol. 28, No. 5, 2005, pp. 381–401.

31. Fricker, J. D. The Integrated Land Use and Statewide Travel Demand Model for Indiana. No. December, 2007.

32. Handy, S., K. R. Shafizadeh, R. Schneider, and others. *California Smart-Growth Trip Generation Rates Study*. 2013.

33. Grantz, K. H., H. R. Meredith, D. A. T. T. Cummings, C. J. E. Metcalf, B. T. Grenfell, J. R. Giles, S. Mehta, S. Solomon, A. Labrique, N. Kishore, C. O. Buckee, and A. Wesolowski. The Use of Mobile Phone Data to Inform Analysis of COVID-19 Pandemic Epidemiology. *Nature Communications*, Vol. 11, No. 1, 2020, pp. 1–8. https://doi.org/10.1038/s41467-020-18190-5.

34. Erlich, A., D. F. Jung, J. D. Long, and C. McIntosh. The Double-Edged Sword of Mobilizing Citizens via Mobile Phone in Developing Countries. *Development Engineering*, Vol. 3, 2018, pp. 34–46. https://doi.org/10.1016/J.DEVENG.2017.11.001.

35. Li, Q., Y. Zheng, X. Xie, Y. Chen, W. Liu, and W.-Y. Ma. Mining User Similarity Based on Location History. 2008.

36. Khetarpaul, S., R. Chauhan, S. K. Gupta, L. V. Subramaniam, and U. Nambiar. Mining GPS Data to Determine Interesting Locations. 2011.

37. Duffus, L. N., S. Alfa, and A. H. Soliman. The Reliability of Using the Gravity Model for Forecasting Trip Distribution. *Transportation*, Vol. 14, 1987, pp. 175–192.

38. Black, W. R. An Analysis of Gravity Model Distance Exponents. *Transportation*, Vol. 2, No. 3, 1973, pp. 299–312. https://doi.org/10.1007/BF00243358/METRICS.

39. Erlander, S., and N. F. Stewart. *The Gravity Model in Transportation Analysis: Theory and Extensions*. Vsp, 1990.

40. Verma, R., T. Yabe, and S. V. Ukkusuri. Spatiotemporal Contact Density Explains the Disparity of COVID-19 Spread in Urban Neighborhoods. *Scientific Reports*, Vol. 11, No. 1, 2021, pp. 1–11. https://doi.org/10.1038/s41598-021-90483-1.

41. Hooper, K. G. *Trip Generation Handbook*. Institute of Transportation Engineers, 2017.

42. Harris, G., and M. D. Anderson. Using Aggregated Federal Data to Model Freight in a Medium-Size Community. 2011.

43. Reia, S. M., P. S. C. Rao, M. Barthelemy, and S. V Ukkusuri. Spatial Structure of City Population Growth. *Nature Communications*, Vol. 13, No. 1, 2022. https://doi.org/10.1038/s41467-022-33527-y.

44. Zhao, P., H. Hu, L. Zeng, J. Chen, and X. Ye. Revisiting the Gravity Laws of Inter-City Mobility in Megacity Regions. *Science China Earth Sciences*, Vol. 66, No. 2, 2023, pp. 271–281.

45. Yabe, T., S. V. Ukkusuri, and P. S. C. Rao. Mobile Phone Data Reveals the Importance of Pre-Disaster Inter-City Social Ties for Recovery after Hurricane Maria. *Applied Network Science*, Vol. 4, No. 1, 2019, pp. 1–18. https://doi.org/10.1007/s41109-019-0221-5.

46. Blumenstock, J., G. Cadamuro, and R. On. Predicting Poverty and Wealth from Mobile Phone Metadata. *Science*, Vol. 350, No. 6264, 2015, pp. 1073–1076.

47. Wang, Z., S. Y. He, and Y. Leung. Applying Mobile Phone Data to Travel Behaviour Research: A Literature Review. *Travel Behaviour and Society*, Vol. 11, 2018, pp. 141–155. https://doi.org/10.1016/j.tbs.2017.02.005.

48. Avery, D. S. *Travel Forecast: 2040 Travel Demands*. Fort Wayne, 2019.